\documentclass{aa}

\usepackage{graphicx}

\def\approxgt{\mathrel{\hbox{\rlap{\lower.55ex \hbox {$\sim$}}
        \kern-.3em \raise.4ex \hbox{$>$}}}}
\def\approxlt{\mathrel{\hbox{\rlap{\lower.55ex \hbox {$\sim$}}
        \kern-.3em \raise.4ex \hbox{$<$}}}}

\begin{document}
\thesaurus{03(11.01.2; 11.09.1: NGC~4945; 11.14.1; 11.19.2; 13.25.2)}

\title{A broad-band X-ray view of NGC~4945}

\author{M. Guainazzi\inst{1,2}, G. Matt\inst{3}, W.N. Brandt\inst{4}, L.A. Antonelli\inst{5}, P. Barr \inst{1}, L. Bassani\inst{6}}

\institute{
{Astrophysics Division, Space Science Department of ESA, ESTEC, Postbus 299,
2200 AG Noordwijk, The Netherlands}
\and
{XMM SOC. VILSPA, ESA, Apartado 50727, E-28080 Madrid, Spain}
\and
{Dipartimento di Fisica, Universit\`a degli Studi ``Roma Tre'', Via della Vasca Navale 84, I-00046 Roma, Italy}
\and
{Department of Astronomy and Astrophysics, The Pennsylvania State University, 525 Davey Lab, University Park, PA 16802, U.S.A.}
\and
{Osservatorio Astronomico di Roma, Via dell'Osservatorio, I-00044 Monteporzio Catone, Italy}
\and
{Istituto Tecnologie e Studio delle Radiazioni Extraterrestri, CNR, Via Gobetti 101, I-40129 Bologna, Italy}
}
   
\offprints{M.Guainazzi}

\date{Received  ; accepted }

\maketitle

\markboth{M.Guainazzi et al.}{A broadband X-ray view of NGC~4945}

\begin{abstract}

We present the results of a BeppoSAX observation of the nearby spiral
galaxy NGC~4945 in the 0.1--200~keV energy band. The nuclear X-ray
emission emerges above $\simeq$7~keV, through an absorber with
column density ${\rm N_H \sim}$ a few $10^{24}$~cm$^{-2}$. Its
remarkable variability (doubling/halving timescale $\sim$10$^4$~s)
is not associated with any appreciable spectral changes, ruling out
variations of the interposed absorber or changes in the primary
continuum shape. The intense iron K$_{\alpha}$ fluorescent emission
line is likely to be produced by the same nuclear absorbing matte.
Our estimate
of the intrinsic nuclear power suggests that the active nucleus
dominates the bolometric energy output ofNGC~4945, if its
nucleus has a typical
quasar ${\rm L_X/L_{bol}}$ ratio. This supports the idea that
X-rays are the best energy band
to search for absorbed AGN.
The 0.1--5~keV emission is extended along the
plane of the galaxy, and most likely due to
a population of unresolved binaries, as believed to happen
in several early-type galaxies. Alternatively, hot gas associated with a
starburst-driven superwind outflow could substantially contribute to the
extended emission above 1~keV.

\end{abstract}


\section{Introduction}

NGC~4945 is a nearby (z=0.0019 or 3.7 Mpc; Mauersberger et al. 1996),
almost edge-on (inclination angle ${\rm \sim 80^{\circ}}$) spiral
galaxy, which exhibits a prominent
dust lane crossing its plane. It is the
third brightest extragalactic
source in the IRAS point source catalog, and most
of its infrared emission is concentrated in a compact nuclear region
(Rice et al. 1988; Brock et al. 1988). It shows both starburst emission
(Heckman et al. 1990; Koornneff 1993) and an H$_2$O megamaser (Dos
Santos \& Lepine 1979). A region of extended optical-line emitting
gas protrudes from the nucleus along the galaxy minor axis
(Nakai 1989; Heckman et al. 1990). It is associated
with a cavity open by a starburst superwind (Moorwood et al. 1996).
X-ray observations with Ginga (Iwasawa et al. 1993) and ASCA (Iwasawa
1997) unveiled  a highly obscured, strongly variable
X-ray source. Actually NGC~4945 turned out to be the brightest
Seyfert~2  and the second brightest
radio-quiet Active Galactic Nucleus (AGN)
after NGC~4151 of the 100~keV sky (Done et al. 1996).

NGC~4945 is one of the best examples of an active galaxy with a composite
nature (starburst plus AGN). At wavelengths shorter than 1~keV,
all its observational properties
can be accounted for by starburst activity alone,
despite its classification as a Seyfert in the V\'eron-Cetty and V\'eron
(1989) catalog. For instance, a
mass-to-light ratio of 0.18 is consistent with the parameter space solely
occupied by starbursters (Oliva et al. 1999).
Also in the Genzel et al. (1998) ISO diagnostic planes NGC~4945 is
located in the regions occupied by starbursts.
However, the presence
of a strong continuum radio source and studies of the off-nuclear optical
spectra have suggested the presence of a LINER
(Whiteoak \& Gardner 1979; Moorwood et al. 1996). The X-ray
observations provided the final proof for the presence of an
active nucleus.
This confirms that hard
X-ray emission can be the most efficient wavelength to identify
the nature of obscured AGN (see {\it e.g.}
the discussion in Vignati et al. 1999 for the case of NGC~6240).

Previous observations in the soft (0.1--2~keV) and
intermediate (2--10~keV) X-rays showed a complexity, which is
far from being completely resolved. The 3--10~keV Ginga spectrum (Iwasawa et
al. 1993) could be fitted with a simple power-law
with a photon index of about 1.7, interpreted as
scattering of the primary nuclear continuum. The same
description fits also the non-simultaneous {\it Ginga},
ASCA and OSSE spectra (Done et al. 1996). A huge (Equivalent
Width, EW, $\sim$1--1.5~keV) iron line has been observed
as well, whose centroid energy is consistent with K$_{\alpha}$
fluorescence from neutral or mildly ionized iron. The ASCA
image is clearly extended, but the irregular, broad shape of
the instrumental Point Spread Function (PSF) has prevented
any detailed studies.
The ROSAT PSPC and HRI images of NGC~4945 revealed a complex
pattern. Brandt et al. (1996)
detected at least five discrete sources, one of them
variable by about one
order of magnitude on timescales of hundreds days.
A diffuse emission up to a distance of about 5$\arcmin$ from the
nucleus, and elongated along the plane of the galaxy, 
was measured as well.

The scientific payload on board BeppoSAX (Boella et al. 1997a) is
particularly well suited to study this source and resolve
at least part of this complexity. It combines the widest X-ray
broadband coverage, the highest sensitivity above 10~keV and the sharpest
2--10~keV PSF ever flown before the advent of {\it Chandra}.
We report in this paper the results of
a BeppoSAX observation performed on July 1999. At the distance
of NGC~4945, 1$\arcmin$ corresponds to about 1.1~kpc.

In this paper:
energies are quoted in the source rest frame; uncertainties
on the spectral parameters are quoted at the 90\% level for one interesting
parameter (${\rm \Delta \chi^2 = 2.71}$); the cosmology used assumes
${\rm H_0 = 50}$~km~s$^{-1}$~Mpc$^{-1}$ and ${\rm q_0 = 0.5}$;
J2000 coordinates are used,
unless otherwise specified. The Galactic column density along the
line of sight is assumed to be ${\rm 1.57 \times 10^{21}}$~cm$^{-2}$
(Heiles \& Cleary 1979). The errors on the energies measured by the
MECS take into account a systematics of 0.8\% at 6~keV
(Guainazzi \& Molendi, 1999). {\sc Xspec 10} was used for spectral
analysis.

The paper is organized as follows.
In Sect.~2 the data reduction procedures are described.
In Sect.~3 we will deal with the analysis of the
LECS and MECS images. We will report the detection in the intermediate
X-rays of two discrete sources in the SE outskirts of the NGC~4945 
spatial profile
(Sect.~3.1), and study the residual unresolved diffuse emission
(Sect.~3.3).
We present also a reanalysis of ROSAT/PSPC archival data of the NGC~4945
field (Sect.~3.2).
The properties of the hard
({\it i.e.}: $>$10~keV) X-ray emission will be studied in
Sect.~4.1. In Sect.~4.2 we will
describe the broadband X-ray spectrum with standard fitting techniques.
We will discuss our findings in Sect.~5 and summarize them in Sect.~6.

\section{Observation and Data Reduction}

NGC~4945 was observed by BeppoSAX
from July 1 04:43 UTC to July 3 1999
09:18 UTC. The scientific payload onboard BeppoSAX comprises two
gas scintillation proportional counters with imaging capabilities:
the Low Energy Concentrator Spectrometer (LECS; Parmar et al.
1997), and the Medium Energy Concentrator Spectrometer
(MECS; Boella et al. 1997b). The LECS has a nominal 
bandpass of 0.1--10~keV, with an energy resolution of $\simeq$4\% at 1~keV
and $\simeq$8\% at 6~keV. The field of view has a diameter of $\simeq$37$\arcmin$.
The angular resolution at 2~keV is 2$\arcmin$.1 Full Width Half Maximum
(FWHM), but degrades to 9$\arcmin$.7 below the Carbon edge ({\it i.e.},
$\approxlt 0.4$~keV). The MECS has similar energy resolution,
but a wider field of view (57$\arcmin$ diameter), a narrower sensitive
bandpass (1.8--10.5~keV) and about a factor of two higher
effective area in the overlapping energy interval,
after the failure of one of the three original units in May 1997.
The 80\% power radius of the PSF
is comprised between 2$\arcmin$.5 and 2$\arcmin$.75.
The BeppoSAX payload also includes
two collimated instruments, mounted on a rocking system
to achieve a continuous monitoring of the background, with a
duty-cycle of 96~s.
The High Pressure Gas Scintillator Proportional Counter
was switched off during the NGC~4945 observation. The
Phoswitch Detector System (PDS; Frontera et al. 1997) possesses an
unprecedented sensitivity in its 13--200~keV nominal bandpass.

The data reduction followed standard procedures, as described
{\it e.g.} in Guainazzi et al. (1999b). In particular,
good time intervals for
the extraction of scientific products were selected
only when the star tracker aligned with the
pointing direction was operative. This condition allows the
best attitude reconstruction. The absolute pointing accuracy under
these conditions is better than 1$\arcmin$, the relative
better than 30$\arcsec$ for sources as bright as NGC~4945. The resulting
net exposure times are about 38.8~ks, 86.7~ks and 87.6~ks in the
LECS, MECS and PDS, respectively.  PDS data
have been reduced using fixed Rise Time thresholds to reject
the particle background, as appropriate for sources brighter
then $0.5$~s$^{-1}$. The PDS points
continuously at the nominal target with only two of its four units,
the others monitoring the background.

MECS background spectra were
extracted from blank deep field exposures, accumulated
by the BeppoSAX Science Data Center (SDC) in the first three years of
the mission. The same method applied to the LECS
spectra yields systematically negative counts below the carbon edge.
We have therefore extracted LECS background spectra from two
semi-annuli in the LECS field of view, and suitably renormalized them
to the source extraction region, as described in
Parmar et al. (1999). The background subtraction in PDS light
curves and spectra has been performed by plain subtraction of the
``off-source'' from the ``on-source'' products. The systematic
uncertainties of this method are lower than $\simeq 0.03$~s$^{-1}$ in
the full PDS sensitive energy bandpass (Guainazzi \& Matteuzzi 1997).
Spectra and light curves of NGC~4945, unless otherwise specified,
have been extracted from circular regions of 8$\arcmin$ radius around the
best-fit centroid of the 2--10~keV image. The position of the
centroid is slightly dependent on energy. However, the dynamical
range of these fluctuations is of the order of 20$\arcsec$, well within
the accuracy of the positional reconstruction. The spectra were
rebinned in order to: a) oversample the FWHM of the energy
resolution by a factor at least 3; b) have at least 20 counts per
spectral channel, to ensure the applicability of $\chi^2$ statistics.
The background-subtracted count rates are: $(2.51 \pm 0.10) \times
10^{-2}$~s$^{-1}$ (0.1--4~keV); $(6.30 \pm 0.10) \times
10^{-2}$~s$^{-1}$ (1.8--10.5~keV); $2.72 \pm 0.04$~s$^{-1}$
(13--200~keV) in the LECS, MECS and PDS, respectively (in brackets the energy
intervals where each instrument is currently well calibrated).
The September 1997 calibration release is adopted throughout this
paper. No known bright sources are present in the PDS field of
view apart from NGC~4945 itself, and the probability of a serendipitous
source with a flux equal or larger than NGC~4945 is $\approxlt 2.5 \times
10^{-4}$, if the 2--10~keV ASCA LogN-LogS (Cagnoni et al. 1998) is adopted
and the lowest extrapolation of the unabsorbed 2--10~keV flux is
assumed (see Sect.~4.1).

\section{Image Analysis in the 0.1-10~keV Energy Band}

\subsection{The serendipitous sources}

\begin{figure}
\includegraphics*[width=8.5cm,height=9.5cm,angle=-90]{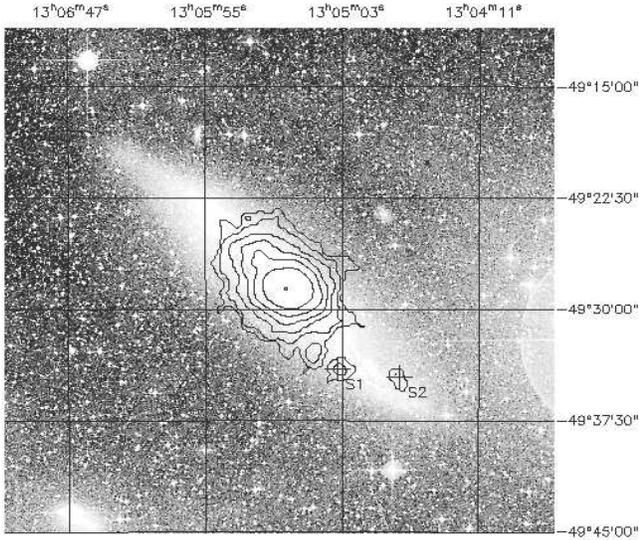}
\caption{Iso-intensity contours for the MECS 2--10~keV image ({\it
solid lines}) superimposed to the ESO DSS optical image. The
contours correspond to steps of 1$\sigma$ of the local background
fluctuations
starting from 4$\sigma$. A Gaussian smoothing with width equal to 2.5 pixels
has been applied to the MECS image for display purposes only. The
best-fit centroid positions of the serendipitous sources S1 and S2
are marked}
\label{fig4}
\end{figure}

\begin{figure}
\includegraphics*[width=7.0cm,height=8.0cm,angle=-90]{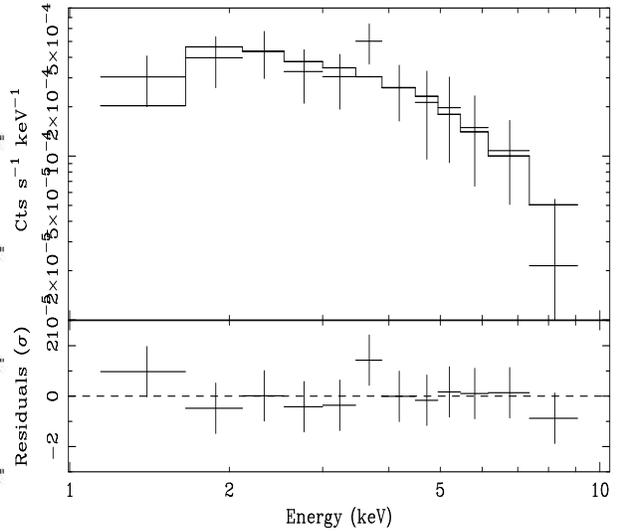}
\caption{Serendipitous source S1
spectrum ({\it upper panel}) and residuals in units of standard
deviations ({\it lower panel}) if a power-law model absorbed by a column
density ${\rm N_H = N_{H,Gal}}$ is applied}
\label{fig11}
\end{figure}

The rather large visual extent of NGC~4945 (3$\arcmin$$\times$20$\arcmin$) is well
encompassed within the fields of view of both the LECS and the MECS.
It is therefore possible with BeppoSAX to compare directly
the shapes of the optical and of the X-ray images. In Fig.~\ref{fig4}
we superpose the 2--10~keV (MECS) iso-intensity contours with the ESO
Digitalized Sky Survey image. The MECS contours are well
aligned with the main plane of the host galaxy.  Actually, part of the
X-ray extension is likely to be due to the presence of discrete sources.
Apart from the nucleus, one source is detected at a signal-to-noise ratio
$>$5 (S1 hereinafter). Its coordinates are 
${\rm \alpha=13^h05^m6.6^s}$; ${\rm \delta=-49^{\circ}33\arcmin 34\arcsec}$,
and it is therefore located 6.2$\arcmin$ SW from the nucleus.
Assuming ${\rm N_H = N_{H,Gal}}$ (the 90\% upper limit
on ${\rm N_H}$ if left free in the fit is
${\rm 1.1 \times 10^{23}}$~cm$^{-2}$), its
spectrum (extracted from a 2$\arcmin$ circular region) can be modeled with a
power-law with ${\rm \Gamma = 1.7 \pm^{0.6}_{0.3}}$
(${\rm \chi^2 = 4.4/9}$~dof; see Fig.~\ref{fig11})
or a thermal bremsstrahlung with
${\rm kT > 4}$~keV (${\rm \chi^2 = 3.8/9}$~dof).
The net background subtracted
count rate is ${\rm (1.7 \pm 0.2) \times 10^{-3}}$~s$^{-1}$ in the 2--10~keV
band, corresponding to a flux of
$\simeq 1.7 \times 10^{-13}$~erg~cm$^{-2}$~s$^{-1}$ and a rest-frame
luminosity of $\simeq 3 \times 10^{38}$~erg~s$^{-1}$
at the distance of NGC~4945. The search for
a periodic modulation has produced no significant peak, hardly
surprising, since only 147 photons are counted from S1.
Another detection at the 4$\sigma$ level
[count rate ${\rm (7.3 \pm 1.7) \times 10^{-4}}$~s$^{-1}$] corresponds to the
following position:
${\rm \alpha=13^h04^m41.9^s}$; ${\rm \delta=-49^{\circ}34\arcmin 22\arcsec}$
(9$\arcmin$ SW from the center). The faintness of this source, and its
location below the MECS strongback rib prevent us from performing
any reliable flux determination or spectral analysis.

\begin{figure}
\includegraphics*[width=8.5cm,height=8.0cm]{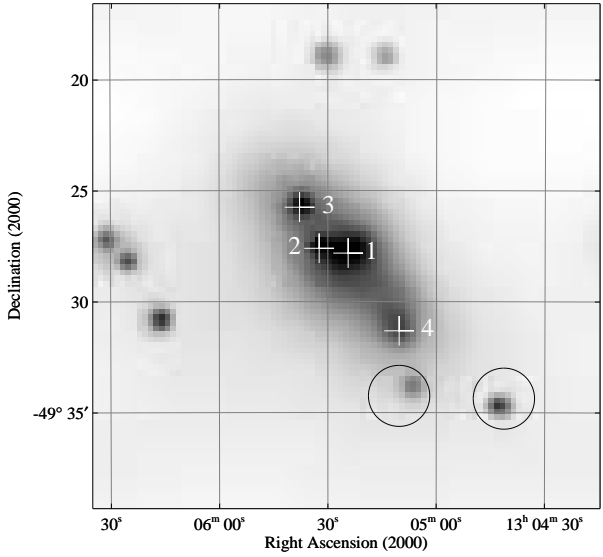}
\caption{
Adaptively smoothed image of the ROSAT/PSPC NGC~4945 field.
The circles indicate the position of S1 ({\it left}) and S2 ({\it right})
and have a radius of 1$\arcmin$, corresponding to the typical MECS error circle. The {\it white crosses} mark the position of the sources listed in
Tab.~\ref{tab5}}
\label{fig13}
\end{figure}

In Fig~\ref{fig13}, we show the NGC~4945 field
imaged by the ROSAT/PSPC on August 1992 and July 1993, for a total
net exposure time of 14.2 and 9.0~ks, respectively.
The BeppoSAX sources coincide with two
of those detected by ROSAT within the 1$\arcmin$ positional accuracy.
The extrapolation of the S1 MECS best-fit into the PSPC band predicts 
$48 \pm 6$ photons, reasonably close to the observed 33
PSPC background-subtracted source counts.
Despite the relatively high Galactic latitude of NGC~4945 (${\rm b =
13.340}$),
the possibility that either or both S1 and S2 is a Galactic
foreground object cannot be ruled out by our data alone. No known source
is included in the {\sc Simbad} or {\sc Ned} databases within 2$\arcmin$
from the best-fit MECS positions. If ${\rm N_H = 10^{20}}$~cm$^{-2}$
(appropriate for a Galactic
source at the distance of approximately 0.5~kpc),
the extrapolation of the MECS best-fit would exceed the ROSAT flux by
a factor of four, which could be accounted for by intrinsic
variability. Analogously the extrapolation of the MECS spectrum remains
consistent with the ROSAT detection within a factor of a few for absorbing
column density $\approxlt 10^{22}$~cm$^{-2}$. It is therefore also 
possible that S1 is a background object, seen through a
moderate column density of the galaxy.

\subsection{The ROSAT nuclear sources}

The ROSAT PSPC detects four sources within the innermost NGC~4945 5$\arcmin$
(see Fig.~\ref{fig13}). Their coordinates and count rates are listed
in Tab.~\ref{tab5}. All of them exhibit a certain amount of
\begin{table}
\begin{tabular}{lccccc} \hline \hline
& ${\alpha}$ & $\delta$ & ${\rm CR}$$^{a.b}$ & ${\rm CR}$$^{a,c}$ & ${\rm d}$$^d$ \\
& & & (10$^{-2}$~s~$^{-1}$) & (10$^{-2}$s~$^{-1}$) & (${\rm \arcmin}$) \\ \hline
\#1 & $13^h05^m24.4^s$ & $-49^{\circ}27\arcmin 49\arcsec$ & $1.36 \pm 0.11$ & $1.65 \pm 0.15$ & 0.6 \\
\#2 & $13^h05^m32.4^s$ & $-49^{\circ}27\arcmin 36\arcsec$ & $1.19 \pm 0.10$ & $0.83 \pm 0.11$ & 0.9 \\
\#3 & $13^h05^m37.8^s$ & $-49^{\circ}25\arcmin 45\arcsec$ & $0.78 \pm 0.09$ & $1.18 \pm 0.12$ & 2.9 \\
\#4 & $13^h05^m10.3^s$ & $-49^{\circ}31\arcmin 19\arcsec$ & $0.44 \pm 0.07$ & \dots$^e$ & 4.3 \\ \hline \hline
\end{tabular}

\noindent
$^a$0.1--2.4~keV PSPC background subtracted count rate

\noindent
$^b$August 1992 observation

\noindent
$^c$July 1993 observation

\noindent
$^d$distance from the galaxy optical nucleus

\noindent
$^e$not detected

\caption{Properties of the sources detected in the ROSAT/PSPC NGC~4945 field within 5$\arcmin$ from the optical nucleus}
\label{tab5}
\end{table}
variability. It is interesting to notice that none of them strictly coincides
with the optical nucleus of the galaxy, the closest being located
0$\arcmin$.6 SE. An extended diffuse emission along the galactic plane
is also present, whose integrated count rate
is about 3--4$\times 10^{-2}$~s$^{-1}$ ({\it i.e.}, about the same
as the sum of the detected discrete sources). No reliable
spectral information can be obtained from any of the above sources. If
their spectrum is a power-law with ${\rm \Gamma = 2}$ (1), they may contribute
$\simeq$15\% (50\%) of the observed 2--10~keV MECS flux.

\subsection{The diffuse emission in BeppoSAX}

\begin{figure*}
\hbox{
\includegraphics*[width=8.0cm,height=8.0cm]{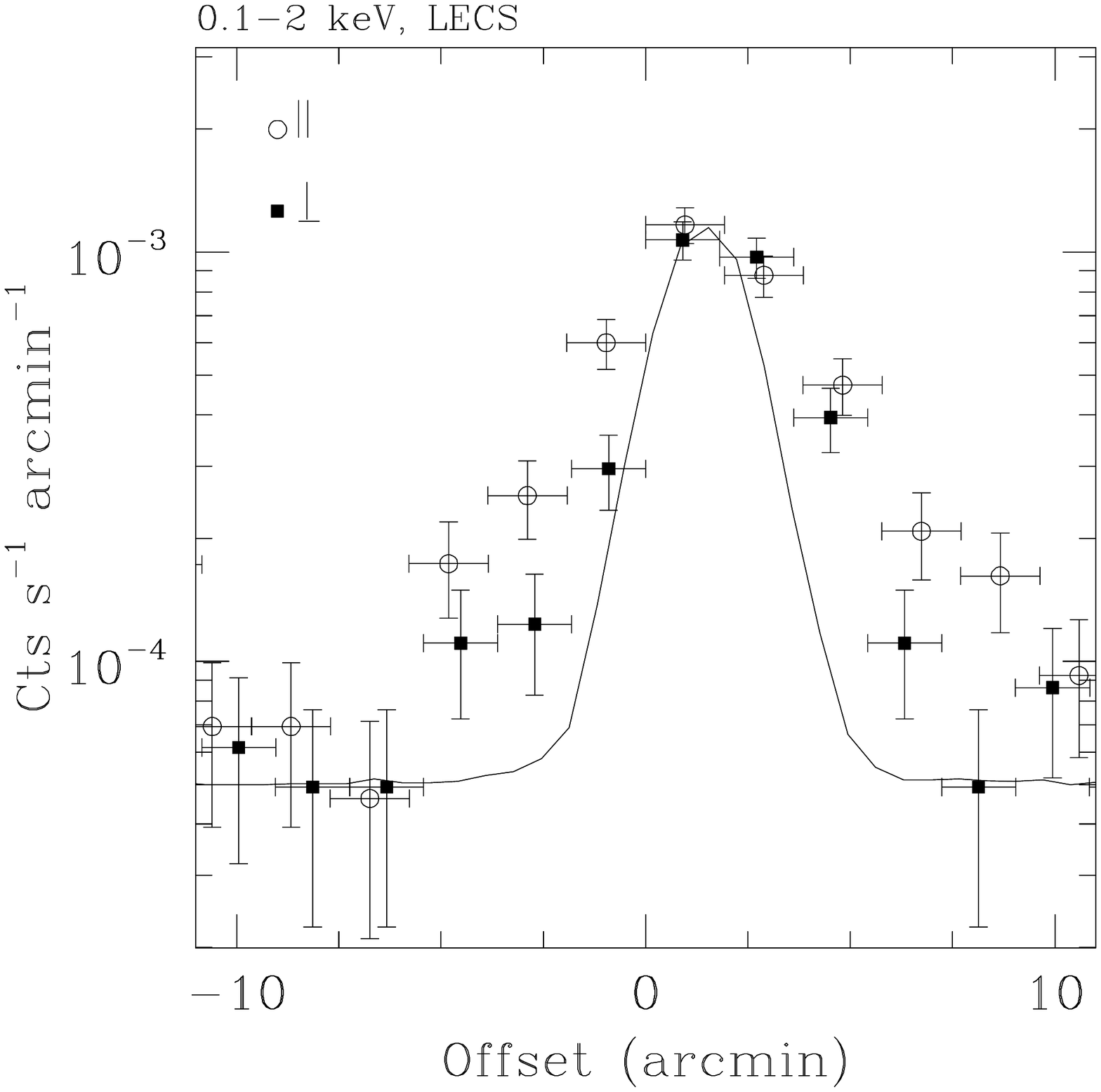}
\hspace{1.0cm}
\includegraphics*[width=8.0cm,height=8.0cm]{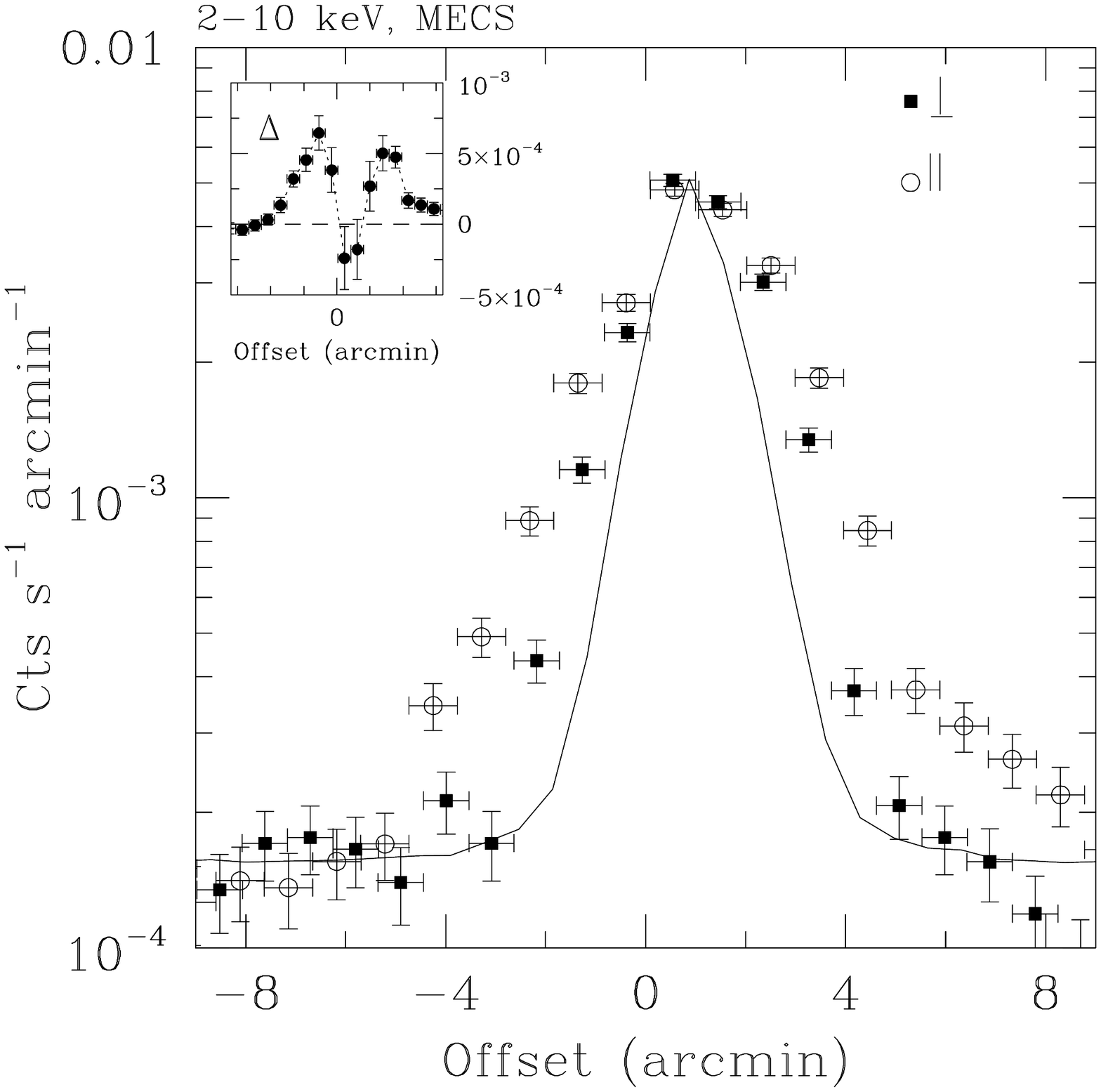}
}
\caption{Count rate profiles along directions parallel ({\it
empty circles}) and perpendicular ({\it filled squares}) to the
optical plane of NGC~4945 for the LECS 0.1--2~keV
({\it left panel}) and the MECS 2--10~keV ({\it right panel}) images.
The {\it solid lines} represent the radial profile for the BeppoSAX
observations of LMC~X-3 and 3C273, respectively.
The offset positive sense is taken
from NE to SW,
therefore in the sense of decreasing RA. In the {\it
inset} of the {\it right panel} the quantity ${\rm \Delta}$ is plotted
as a function of offset angle, defined as the difference between the
observed count rates
in the parallel and perpendicular directions at the same offset distance.
Each tickmark in this {\it inset} corresponds to 2$\arcmin$}
\label{fig5}
\end{figure*}

In order to provide
a quantitative estimate of the intrinsic extension and ellipticity
of the BeppoSAX X-ray images, we extracted the counts along two perpendicular
strips crossing the optical center of the galaxy in the 0.1-2~keV (LECS)
and 2--10~keV (MECS) energy bands,
respectively. The strip directions are defined
by the following pair of coordinates:
(${\rm \alpha=13^h05^m44.3^s}$; ${\rm \delta=-49^{\circ}25\arcmin 22\arcsec}$) and 
(${\rm \alpha=13^h05^m12.3^s}$; ${\rm \delta=-49^{\circ}30\arcmin 34\arcsec}$) for
the strip parallel to
the plane of the galaxy ({\it i.e.}: NE to SW);
(${\rm \alpha=13^h05^m36.1^s}$; ${\rm \delta=-49^{\circ}30\arcmin 26\arcsec}$) and 
(${\rm \alpha=13^h05^m13.1^s}$; ${\rm \delta=-49^{\circ}26\arcmin 10\arcsec}$) for
the strip perpendicular to the plane of the galaxy ({\it i.e.}: NW to SE). The
width of the strips is 4\arcmin, comparable with
the intrinsic width of the instrumental PSF (Parmar et al. 1997;
Boella et al. 1997b). This choice is actually likely to discard
the bulk of the photons emitted below the carbon edge,
where the LECS detector PSF is the broadest.
However, this effect is negligible in our case, because the 
photoelectric absorption due to the matter in our Galaxy
along the line of sight of NGC~4945 is large enough
to substantially suppress the photons in this band. The results are
shown in Fig.~\ref{fig5}, as count rate space density as a
function of the offset angle towards the galaxy center (note
that the offset {\it increases} for {\it decreasing} RA).
The MECS image shows clearly a larger extent
in the direction parallel to the galaxy plane for all offset angles
between -4$\arcmin$ and 8$\arcmin$. Comparison at larger angles is made impossible by
the shadow of the strongback support ring in the MECS field of view.
The difference between the profiles along the parallel and perpendicular
directions is rather symmetric within 4$\arcmin$, while an excess tail is evident
at larger offset radii only for positive offsets
(cf. the inset in the right panel of Fig.~\ref{fig5}), due to
the presence of the discrete sources detected by the MECS.

In the same Fig.~\ref{fig5}, we compare the observed NGC~4945 profile with
point-like sources, whose data have been
reduced under the same experimental conditions as the NGC~4945 ones. In
order to avoid any systematic effects due to the dependence of the PSF
with energy, we have chosen two sources
contained in the archive of the BeppoSAX public observations,
whose spectrum is similar
to that observed in NGC~4945. The 0.1--2~keV LECS spectrum of NGC~4945
can be formally well approximated by a photoelectrically absorbed power-law,
with parameters: ${\rm \Gamma = 1.2 \pm^{0.3}_{0.2}}$; ${\rm N_H =
(1.2 \pm^{1.5}_{0.8}) \times 10^{21}}$~cm$^{-2}$. This model is
consistent, within the statistical uncertainties, with that describing
the LECS spectrum of the black-hole candidate LMC~X-3 in the same
energy band (Siddiqui et al. 1999).
The 2--10~keV MECS continuum of NGC~4945 is well approximated by
a power-law, with ${\rm \Gamma = 1.91 \pm^{0.28}_{0.16}}$.
The comparison for the
MECS is made against the profile of the quasar 3C273 (Grandi et al. 1997).
The presence of a prominent iron line in the NGC~4945 spectrum (Iwasawa
et al. 1993), which is absent in 3C273, makes a negligible difference
in this context.
Even the profile perpendicular
to the galaxy plane is broader than the instrumental PSF within
$\pm$3$\arcmin$--4$\arcmin$. However, caution has to be used in interpreting
the last result, as the residual inaccuracies in the attitude reconstruction
could significantly affect the significance of the last result, whereas no
doubt exists about the true nature of the extension along the galaxy
plane.

\begin{figure}
\includegraphics*[width=8.0cm,height=8.0cm]{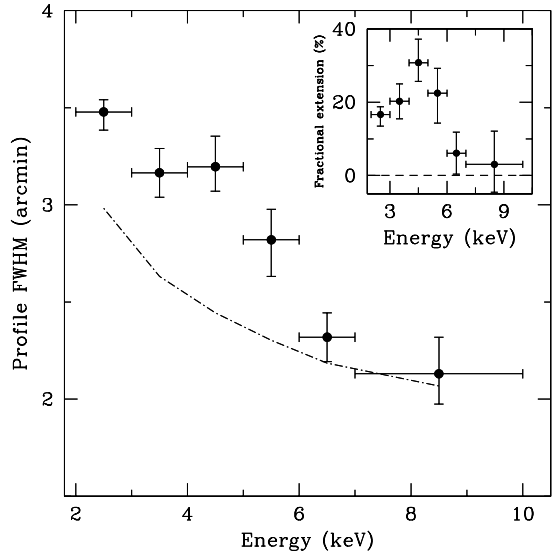}
\caption{FWHM of the X-ray image profile along the host galaxy plane
as a function of energy. The {\it dash-dotted} line marks the values
expected from the on-flight MECS calibration (Boella et al. 1997b;
Chiappetti 1998). In the inset, the percentage fractional difference
between the observed and the expected values as a function of energy is
shown.}
\label{fig10}
\end{figure}

As a further step, we have studied the energy dependence of the X-ray
image profile along the plane of the galaxy with the MECS (an analogous
study with the LECS is hampered by the much poorer statistics and
the broader instrumental PSF). The
results of such a study are shown in Fig.~\ref{fig10}. They
are compared with the width expected on the basis of the instrumental PSF,
weighted on the energy assuming a power-law spectrum of photon index
$\Gamma = 1.8$ (dash-dotted line in Fig.~\ref{fig10}). The X-ray MECS image
is significantly extended in the whole range between 2 and 5~keV, whereas
it becomes consistent with that expected from a point-like source
for ${\rm E \approxgt 6}$~keV. It is difficult with our
data to assess a systematic trend between 2 and 5~keV. If we consider the
face-on values, the profile width has a peak at $\simeq$4~keV.

\section{The X-ray Spectral Energy Distribution}

\begin{figure}
\includegraphics*[width=8.0cm,height=8.0cm]{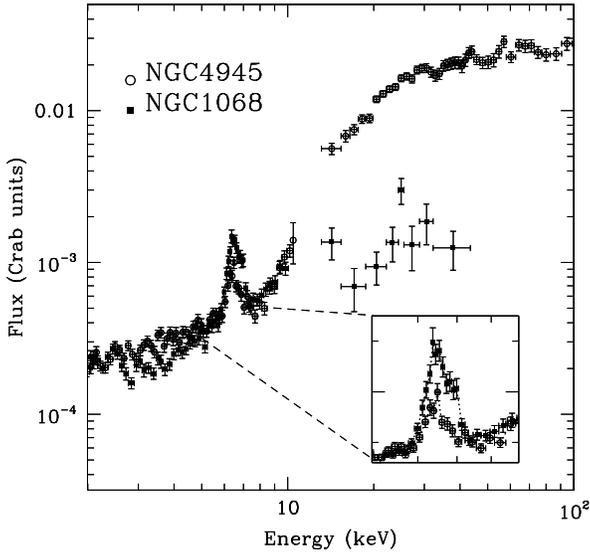}
\caption{Comparison of the MECS and PDS spectra of NGC~1068 ({\it
filled squares}) and NGC~4945 ({\it empty circles}). Both spectra
have been normalized to the Crab spectrum, which is a power-law
with a photon index $\simeq$2. Each data point corresponds to a signal-to-noise
ratio $>$10, except the PDS data points of NGC~1068, for which
the signal-to-noise ratio is $>$3. In the {\it inset} a zoom of the
5--8.5~keV energy band, with the flux in linear scale}
\label{fig8}
\end{figure}

In Fig.~\ref{fig8} we compare the $>$2~keV spectra of NGC~4945 and
NGC~1068, after they have been divided by the spectrum of the Crab Nebula.
Since the latter is a power-law with photon index $\simeq$2, this
technique is very similar to a ${\rm \nu F_{\nu}}$ plot.
In the intermediate X-ray band, the two spectra
are similar. A careful inspection suggests
that the emission line feature around 6~keV is broader and
more intense in NGC~1068, indicating that more ionization stages
contribute to its profile than in NGC~4945. The most remarkable
difference is, however, at energies $>$10~keV. NGC~4945 is more than
one order of magnitude brighter in the PDS, suggesting that a
further component  emerges - actually well connected with the higher end
of the MECS sensitive bandpass. It is straightforward to
identify this component with the primary nuclear continuum transmitted through
an absorber with a column density of a few $10^{24}$~cm$^{-2}$, originally
discovered by Iwasawa et al. (1993). On the other hand, in NGC~1068,
the archetypical example of totally Compton-thick Seyfert~2, the nuclear
emission is completely suppressed.

\subsection{The high-energy emission}

\begin{figure}
\includegraphics*[width=7.0cm,height=8.0cm,angle=-90]{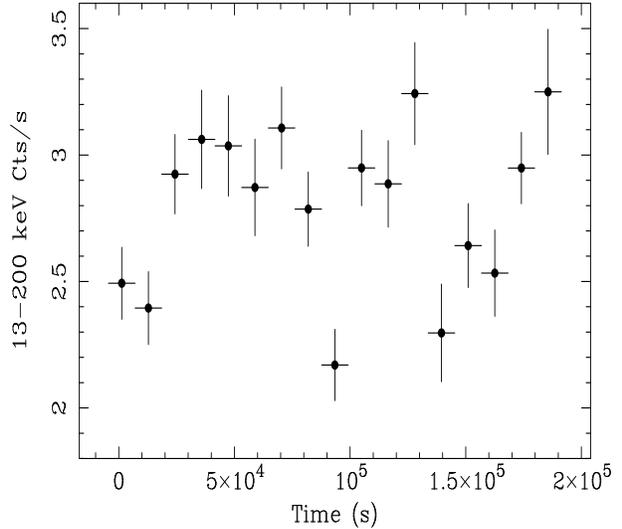}
\caption{Light curve in the 13--200~keV energy band (PDS). The binning
time is 11520~ks, approximately corresponding to two BeppoSAX orbits}
\label{fig1}
\end{figure}

\begin{figure}
\includegraphics*[width=8.0cm,height=8.0cm,angle=-90]{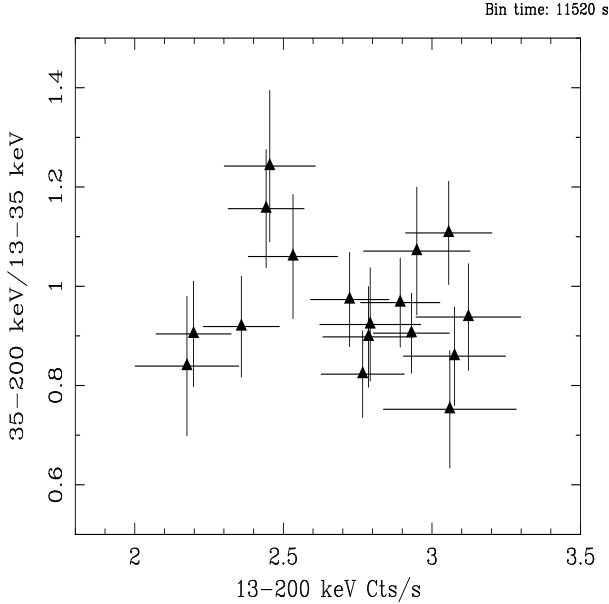}
\caption{Hardness ratio between the count rates in the
35--200~keV and 13--35~keV energy bands, as a function of the
13--200~keV count rate}
\label{fig2}
\end{figure}

As shown in Fig.~\ref{fig1}, the PDS full band light curve exhibits
a remarkable variability. In three events the flux increases by
about 60\% in $\simeq$$3 \times 10^4$~s, whereas a decrease by a
factor $\simeq$30\% in $\approxlt 10^4$~s
represents the most extreme variability episode.
We have searched in vain for spectral changes associated with these
flux variations. In Fig.~\ref{fig2} we show the hardness ratio (HR)
between the count rates in the 13--35 and 35--200~keV energy bands. No
clear deviation from constancy is observed. If we fit the data points
in Fig.~\ref{fig2} with a constant, $\chi^2 = 18.0/17$~degrees of
freedom (dof). We will therefore hereinafter focus on the time-averaged
spectrum only.

\begin{figure}
\includegraphics*[width=8.0cm,height=8.0cm,angle=-90]{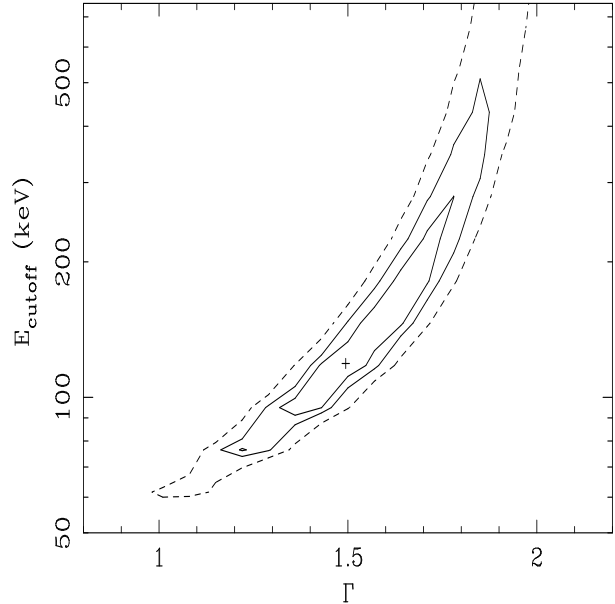}
\caption{Iso-$\chi^2$ confidence contours for the spectral index
versus the high-energy cutoff, for the best-fit model of
the second row of Tab.~1. Contours are at 67\%, 90\% ({\it
solid lines}) and 99\% ({\it dotted lines}) for two interesting
parameters.}
\label{fig3}
\end{figure}

An absorbed power-law model yields a marginally acceptable fit to the
PDS spectrum (${\rm \chi^2 = 62.3/47}$~dof). We first modeled the extinction
as an exponential function of the energy ${\rm \exp^{-N_H \sigma(E)}}$,
where the cross-section ${\rm \sigma(E)}$ includes both the
effects of photoelectric absorption (model {\tt wabs} in {\sc
Xspec}, with the abundances of Morrison \& McCammon 1983)
and of Compton scattering (model {\tt cabs} in {\sc Xspec}).
The inclusion of a high-energy
cutoff improves the quality of the fit by ${\rm \Delta \chi^2 = -7.7}$,
for the decrease of one degree of freedom, which is significant at $>99.2\%$
confidence level
according to the F-test. The best-fit parameters and results
are summarized in Tab.~\ref{tab1}.
\begin{table}
\begin{tabular}{lcccc} \\ \hline \hline
Model & ${\rm N_H}$ & $\Gamma$ & ${\rm E_c}$ & $\chi^2/$dof \\
& ($10^{24}$~cm$^{-2}$) & & (keV) & \\ \hline
PL$^a$ & $5.4 \pm 0.8$ & $2.00 \pm ^{0.08}_{0.07}$ & \dots & 62.3/47 \\
CP$^a$ & $3.9 \pm ^{0.8}_{0.9}$ & $1.5 \pm^{0.3}_{0.4}$ & $110 \pm^{130}_{30}$ & 54.7/46 \\
CP$^b$ & $2.2 \pm^{0.3}_{0.4}$ & $1.4 \pm 0.3$ & $130 \pm^{250}_{60}$ & 57.7/46 \\ \hline \hline
\end{tabular}

\noindent
$^a$Compton absorption after model {\tt cabs} in {\sc Xspec}

\noindent
$^b$Compton scattering after Matt et al. (1999b; absorption model for
optically thick Compton scattering). In comparison to the same model
without high-energy cutoff ${\rm \Delta \chi^2 = -6.9}$

\caption{Best-fit parameters for the fits on the PDS spectrum alone.
PL~=~power-law, CP~=~cutoff power-law. }
\label{tab1}
\end{table}
In Fig.~\ref{fig3} we show the iso-$\chi^2$
confidence contours for ${\Gamma}$
versus the cutoff energy ${\rm E_c}$.
At 90\% confidence level for two interesting parameters, ${\rm E_c}$
is comprised between 70 and 500~keV, the value corresponding
to the best-fit being $\simeq$110~keV.
The addition of a Compton-reflection component from a plane-parallel
infinite edge-on slab to the best-fit model (implying the decrease of
a further degree of freedom to the fit, the relative normalization
between the reflected and the transmitted component R; model
{\tt pexrav} in {\sc Xspec}, Magdziarz \& Zdziarski 1995) does
not yield any further improvements to the fit. The 90\% upper limit on
R is 0.5. The 20--100~keV observed flux is ${\rm (2.89 \pm 0.04) \times
10^{-10}}$~erg~cm$^{-2}$~s$^{-1}$. The extrapolation of the best fit
model (second row of Tab.~\ref{tab1}) to the 2--10~keV (0.1--200~keV)
energy band yields an unabsorbed flux of ${\rm \simeq 3.5 \times 10^{-9}}$
(${\rm 1.83 \times 10^{-8}}$)~erg~cm$^{-2}$~s$^{-1}$, corresponding
to a rest-frame luminosity of ${\rm \simeq 5.3 \times 10^{43}}$
(${\rm 2.8 \times 10^{44}}$)~erg~s$^{-1}$.

At the level of $4 \times 10^{24}$~cm$^{-2}$, the matter is optically
thick to Compton scattering. The {\tt cabs} model in {\sc Xspec}
assumes, on the contrary, the simple Thompson cross-section.
Neglecting the effects of electron scattering in the column may
lead to a substantial overestimate of the true column density and,
consequently, of the intrinsic nuclear luminosity (Leahy et al. 1989).
Matt et al. (1999b) have recently developed a self-consistent model
for X-ray absorption by matter which is optically thick to
Compton-scattering and has a large covering factor.
The application of such a model yields
about 40\% lower value for the column density. Moreover, the extrapolated
intrinsic luminosities differ significantly. In the Matt et al. (1999b)
model a fraction of the photons are scattered into the line of sight,
and the nuclear luminosity required to explain the observed flux
is lower. The extrapolation of the best fit model (row 3 in Tab.~\ref{tab1})
yields an unabsorbed 1--10~keV (0.1--200~keV) flux of ${\rm \simeq 2.5
\times 10^{-10}}$ (${\rm 1.14 \times 10^{-9}}$)~erg~cm$^{-2}$~s$^{-1}$,
corresponding to a rest frame luminosity of ${\rm \simeq 3.8 \times
10^{42}}$ (${\rm 1.77 \times 10^{43}}$)~erg~s$^{-1}$. We will refer to
these values hereinafter.

\subsection{The broadband 0.1-200~keV spectrum}

Given the spatial and spectral complexity emerging from the
above analysis, a global description of the 0.1--200~keV
spectrum in terms of standard fitting technique at the moderate spatial
resolution and sensitivity provided by the BeppoSAX instruments is hardly
more than an academic task. The soft and intermediate X-ray emission is
likely to be produced by the superposition of radiation scattered/reprocessed
in the nuclear environment (probably most contributing to the huge iron line),
a population of discrete unresolved sources (binaries, cataclysmic variables,
supernova remnants) and truly diffuse gas emission, each possibly seen
through absorbers with different density and/or physical conditions.
Such complex scenario will be fully understood when future observation with
arc-second spatial resolution and/or much better sensitivity will be
available. Given all these caveats, in this Section we will, however,
try to accomplish this
task, whose results has allowed us to extend previous studies of this kind.

\begin{figure}
\includegraphics*[width=7.5cm,height=8.0cm,angle=-90]{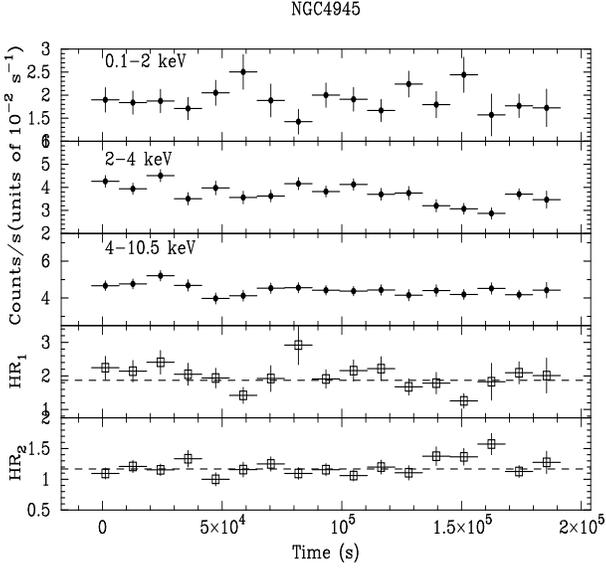}
\caption{Light curves in the 0.1--2~keV, 2--4~keV and 4--10.5~keV
energy bands, and hardness ratios between adjacent bands. The
binning time is 11520~s. HR$_1$ is defined as the ratio between the
counts in the 2--4~keV and 0.1--2~keV energy bands. HR$_2$ is defined
as the ratio between the
counts in the 4-10.5~keV and 2--4~keV energy bands.}
\label{fig9}
\end{figure}

In Fig.~\ref{fig9} we show the light curves in the 0.1--2~keV, 2--4~keV
and 4--10.5~keV energy bands, along with the corresponding HRs
between adjacent bands. No significant variability of either quantity
is observed.
This is of course not surprising, because the bulk of the emission
below 6~keV is extended on several kpc scale (see Sect.~3.3).
Fit with a constant line of the HRs yield $\chi^2$ of
22.3/17~dof and 17.9/17, for HR$_1$ ($\equiv$2--4~keV/0.1--2~keV)
and HR$_2$ ($\equiv$10.5--4~keV/2--4~keV), respectively. We are therefore
justified in focusing on the time-averaged spectrum only.

The broadband spectrum of NGC~4945 was fitted using the following general
formula:
$$
F(E) = \exp^{-[\sigma_{ph} N_{H,Gal}]} \{ \exp^{-[\sigma_{ph}(E) N_{H,1}]} A(E)
$$
$$
+\exp^{-[\sigma_{ph}(E) N_{H,2}]} B(E) + G(E)
$$
$$
+ \exp^{-\{ [\sigma_{ph}(E) + \sigma_{C}(E)] N_{H,tr} \}} \exp^{(-E/E_c)}
NE^{-\Gamma_{tr}} \}
$$
The last term in brackets is the nuclear transmitted component emerging in
the PDS, with normalization ${\rm N}$.
${\rm \sigma_{ph}(E)}$ and ${\rm \sigma_{C}(E)}$ are the
photoelectric and Compton cross-sections, respectively. We will use
the {\tt cabs} {\sc Xspec} implementation for the latter.
${\rm A(E)}$ and ${\rm B(E)}$ are two continuum models, chosen
according to one of either the following scenarios:
\begin{itemize}
\item[a)] 
${\rm A(E)}$ is
a power-law, produced by the scattering of the primary continuum by ionized
matter, and ${\rm B(E)}$ is
the thermal emission from a collisionally excited plasma.
The index of the former is tied to be the same as ${\rm \Gamma_{tr}}$. We will
use the {\tt mekal} {\sc Xspec} implementation for the
latter throughout this paper.
This model provides a reasonable description of the spectra of
some reflection-dominated Seyfert~2 galaxies, like Circinus Galaxy, NGC~1068
(Guainazzi et al. 1999a), or NGC~6240 (Iwasawa \& Comastri 1998; Vignati et al.
1999). All these objects are characterized by a strong circumnuclear starburst
and/or are infrared ultraluminous galaxies
\item[b)] ${\rm A(E)}$ is
a thermal plasma emission and ${\rm B(E)}$ is either
a blackbody (b1) or a second thermal
plasma (b2), which is
the template model successfully employed to fit the {\it Einstein},
ROSAT and ASCA bulge emission of early-type galaxies (Bregman et al.
1995; Matsumoto et al. 1997; Kim et al. 1996; Irwin \& Bregman 1999) 
\item[c)]  ${\rm A(E)}$ is
a power-law and ${\rm B(E)}$ is
a Compton-reflection, both with the photon
index tied to be the same as ${\rm \Gamma_{tr}}$. This
scenario follows the idea that the primary nuclear continuum undergoes
both scattering by a warm plasma and Compton reflection,
along two different optical
paths.
\end{itemize}

To all models a Gaussian emission line ${\rm G(E)}$ was added. We allowed
different absorbing column densities for the nuclear (warm-scattered
and Compton-reflected power-law) and the thermal/non-nuclear components.
Finally, we included in the models
a power-law, with all the parameters
frozen to the best-fit values obtained fitting the S1 spectrum alone
(this contribution is not explicity indicated in the above formula
for sake of clarity).

\begin{figure*}
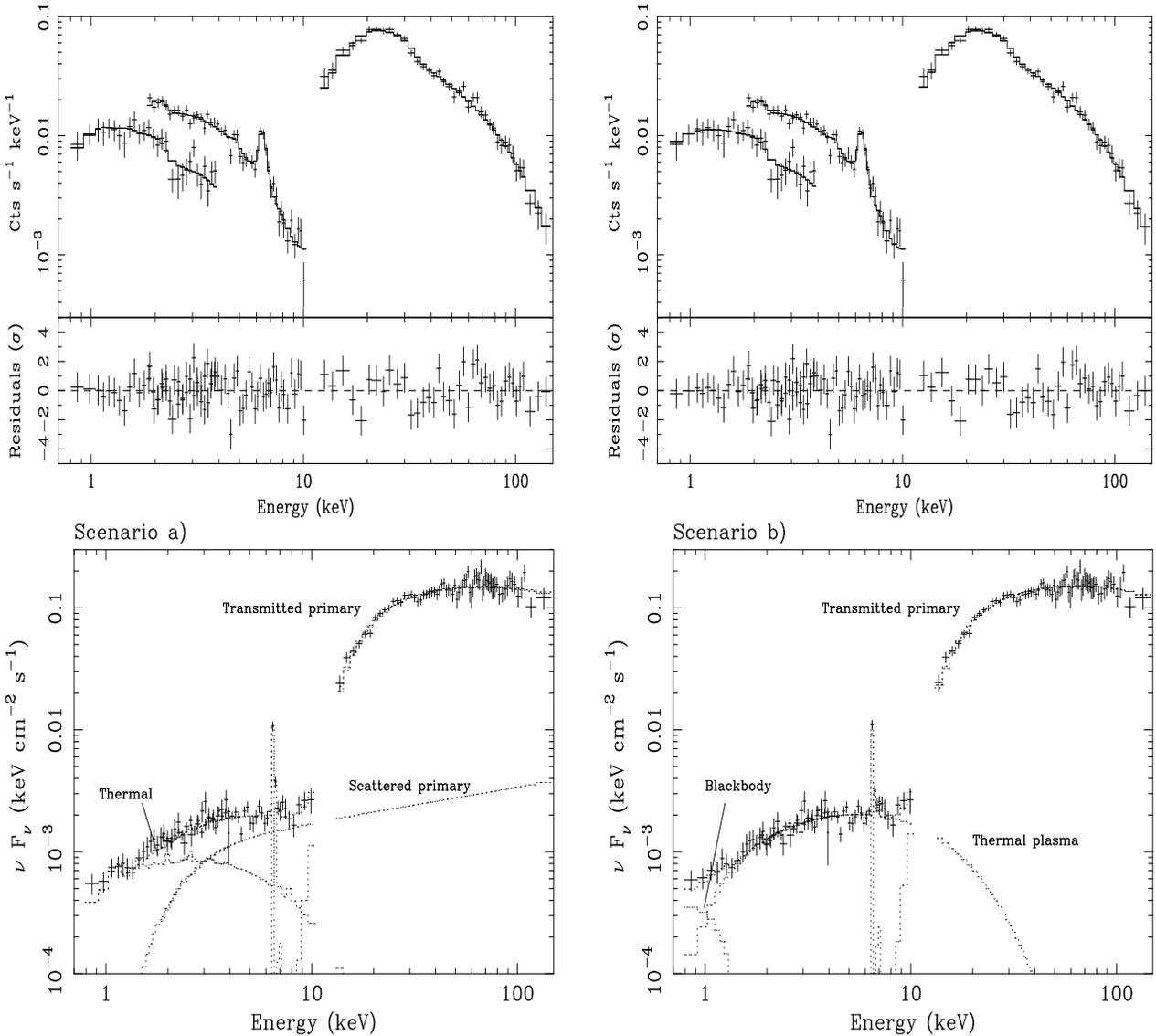

\hbox{
\includegraphics*[width=7.5cm,height=8.0cm,angle=-90]{fig12a.ps}
\hspace{0.5cm}
\includegraphics*[width=7.5cm,height=8.0cm,angle=-90]{fig12b.ps}
}
\hbox{
\includegraphics*[width=7.5cm,height=8.0cm,angle=-90]{fig12c.ps}
\hspace{0.5cm}
\includegraphics*[width=7.5cm,height=8.0cm,angle=-90]{fig12d.ps}
}
\caption{Spectra and residuals in units of standard deviations
({\it upper panels}) and corresponding best-fit Spectral Energy
Distributions {(\it lower panels}) for scenarios a) ({\it left}) and
b1) ({\it right})}
\label{fig12}
\end{figure*}

The best-fit parameters and results are shown in Tab.~\ref{tab3}. The fit
\begin{table*}
\begin{footnotesize}
\begin{center}
\begin{tabular}{lccccccccccc} \hline \hline
& \multicolumn{4}{c}{Thermal components} & \multicolumn{5}{c}{Power-law} & \\
& ${\rm N_{H,1}}$ & ${\rm kT}$ & ${\rm Z}$ & ${\rm kT}$ & ${\rm N_{H,2}}$ & $\Gamma$ & ${\rm f_s^a/R}$ & ${\rm N_{H,tr}}$ & ${\rm E_{cutoff}}$ & $\chi^2$ \\ 
& ($10^{21}$~cm$^{-2}$) & (keV) & & (keV) & ($10^{21}$~cm$^{-2}$) & & (\%) & ($10^{24}$~cm$^{-2}$) & (keV)  & \\ \hline
a) & $<$3.4 & $3 \pm^6_2$ & $0.2 \pm^{0.6}_{0.2}$ & \dots & $26 \pm 14$ & $1.61 \pm 0.16$ & 0.4/$<$1$^b$ & $4.5 \pm 0.7$ & $190 \pm^{100}_{90}$ & 158.3/127 \\
b1)$^c$ & $6 \pm 3$ & $8 \pm 2$ & $0.14 \pm^{0.23}_{0.14}$ & $0.11 \pm^{0.03}_{0.04}$ & \dots & $1.59 \pm^{0.17}_{0.37}$ & 0.4/$<$12$^b$ & $4.5 \pm 0.7$ & $140 \pm^{100}_{50}$ & 158.3/127 \\
b2)$^c$ & $7 \pm 4$ & $8 \pm 2$ & $0.12 \pm^{0.27}_{0.12}$ & $0.26 \pm_{0.11}^{0.23}$ & \dots & $1.6 \pm^{0.2}_{0.3}$ & \dots/$<$12$^b$ & $4.4 \pm^{0.8}_{0.6}$ & $140 \pm^{170}_{40}$ & 155.3/125 \\
c) & \dots & \dots & \dots & \dots & $1.8 \pm^{1.6}_{0.9}$ & $1.70 \pm^{0.11}_{0.05}$ & 0.4/$<$2 & $5.1 \pm^{0.5}_{0.3}$ & $170 \pm^{110}_{30}$ & 172.2/128 \\ \hline \hline
\end{tabular}
\end{center}
\end{footnotesize}

\noindent
$^a$warm scattering fraction

\noindent
$^b$upper limit on the Compton reflection relative normalization parameter. It is calculated after adding this component to the best-fit model

\noindent
$^c$in scenario b1) an optically thin plasma thermal emission (columns 2--4) and a blackbody (column 5) are employed. In scenario b2) the blackbody is substituted by an optically thin plasma thermal emission

\caption{Best-fit parameters and results for the continuum components
in the broadband fits of the
BeppoSAX NGC~4945 observation}
\label{tab3}
\end{table*}
is actually only marginally acceptable for scenarios a) and b). However,
no significant improvement is obtained if a further
component (either a continuum, an emission line or an absorption edge)
is added to any of the above models.
No improvement in the quality of the fit in scenario a) is
obtained if the warm scattered power-law spectral index is untied from 
${\rm \Gamma_{tr}}$. No systematics trend can be recognized in the residuals,
as shown in Fig.~\ref{fig12}. We checked that the rather high $\chi^2$ is
not due to residual systematics effects. 

The fit results strongly argue against any Compton reflection component to
substantially contribute to the broad band spectrum. The scenario
c), which includes it explicitly, yields a significantly worse $\chi^2$
with a comparable number of degrees of freedom. The addition of
a Compton reflection component to the other scenarios is not statistically
justified. In scenario a), its 2--10~keV flux upper limit is
${\rm 5.9 \times 10^{-13}}$~erg~cm$^{-2}$~s$^{-1}$, against a
${\rm 3.1 \times 10^{-12}}$~erg~cm$^{-2}$~s$^{-1}$ flux of the warm
scattered component in the same energy band.

The properties of the iron
line are not strongly dependent on the adopted continuum
(see Tab.~\ref{tab4}). It remains narrow (intrinsic width
${\rm \sigma < 150}$~eV) and
\begin{table}
\begin{center}
\begin{tabular}{lccc} \hline \hline
Model & ${\rm E}$ & ${\rm \sigma}$ & ${\rm I}$ \\
& (keV) & (eV) & (${\rm 10^{-5}}$~photons~cm$^{-2}$~s$^{-1}$) \\ \hline
a) & $6.46 \pm 0.07$ & $<$150 & $4.3 \pm^{0.8}_{1.3}$ \\
b1) & $6.46 \pm 0.07$ & $<$140 & $4.2 \pm 1.1$ \\
c) & $6.48 \pm 0.07$ & $<$140 & $4.2 \pm 0.7$ \\ \hline \hline
\end{tabular}
\end{center}
\caption{Iron line properties for the best-fit models of Tab.~3}
\label{tab4}
\end{table}
consistent with K$_{\alpha}$ fluorescence from neutral or mildly ionized
iron (Fe$<${\sc xx}).
If we extrapolate the best-fit model of raw~3 Tab.~\ref{tab1},
and compare it with the observed line intensity, the iron line
EW against the transmitted nuclear continuum alone is $\simeq$1.3~keV.

Assuming the best-fit scenario b),
the observed flux in the 0.1--2~keV (2--10~keV) energy
band is $1.3 \times 10^{-12}$ ($5.4 \times 10^{-12}$)~erg~cm$^{-2}$~s$^{-1}$,
corresponding to a rest frame unabsorbed luminosity of $1.7 \times 10^{40}$
($8.4 \times 10^{40}$)~erg~s$^{-1}$.

\section{Discussion}

\subsection{The hard X-ray emission}

NGC~4945 is confirmed to be one of the brightest
extragalactic objects above 10~keV, where
we are seeing the primary nuclear continuum, transmitted through
an absorbing column density of a few $10^{24}$~cm$^{-2}$. This
interpretation is confirmed by the detection of rapid variability
in the PDS emission, with a extrapolated doubling/halving time scale
${\rm \tau}$
$\sim 3$--$5 \times 10^4$~s.
Variation of a comparable amount were observed by {\it Ginga}
in the 9.1--30~keV energy band (Iwasawa et al. 1993).
${\rm \tau}$ is actually only an upper limit to
the variability timescales of the primary continuum.
If the absorbing matter is spherically distributed, such a 
relatively rapid variability must be related only to the fraction
of nuclear photons, which are transmitted without being scattered.
For ${\rm \log(N_{H,tr}) = 24.5}$, more than 2/3 of the 20--200~keV
incoming photons are scattered (Matt et al. 1999b). This implies
that the observed variability could be substantially diluted.
If we suppose that the light curve in Fig.~\ref{fig1} is given
by the sum of a constant plateau
(scattered photons) and a variable contribution (unscattered photons), the
latter has 2.6 times
higher dynamical range, or,
conversely, a ${\rm \tau}$ lower by this amount. ${\rm \tau \sim 10^4}$~s
is still not exceptional among Seyfert galaxies. It is interesting to
notice that NGC~4945 is one of the nearby galaxies with the smallest
estimate of the central dark object mass (${\rm M_{DO} \sim 1.6 \times
10^6 M_{\odot}}$; Greenhill et al. 1997). This suggests that NGC~4945
might be an absorbed version of NGC~4051 (Lawrence et al. 1985),
accreting at a sizeable fraction ($\sim 10^{-1}$) of the Eddington rate.
Alternatively, our estimate of the
intrinsic nuclear variability dynamical range could be lowered if
the covering fraction of the absorber is much less than unity. The lack
of any significant detection of Compton reflection in the
broadband X-ray spectrum supports this hypothesis.

It is remarkable that no
spectral variability is associated with these flux changes. This
rules out that changes in the interposing absorbing medium are
responsible for the observed flux variations. On the other hand, this suggests
that the shape of the primary continuum in NGC~4945 is not significantly
dependent on the intensity state. The infrared luminosity,
(${\rm 3 \times 10^{43}}$~erg~s$^{-1}$
if calculated according to Mulchaey et al. 1994), 
and our extrapolation  of the 2--10~keV nuclear intrinsic
luminosity (${\rm 3 \times 10^{42}}$~erg~s$^{-1}$)
lie well on the low luminosity
end of the
correlation observed in Seyferts for these two quantities (Mulchaey
et al. 1994). If ${\rm L_{1-10 \ keV}/L_{bol} \sim 0.05}$,
as typical for quasars with ${\rm L_{1-10 \ keV} < 10^{45}}$~erg~s$^{-1}$
(Elvis et al. 1994), the bolometric luminosity associated to the
AGN is of the same order of magnitude as the observed infrared luminosity,
and hence the AGN dominates the energy output,
at variance with the deductions of Genzel et al. (1998) on the basis of
the ISO spectrum (see Marconi et al. 1999 for a possible explanation of
this discrepancy).

A high-energy cut-off in the primary nuclear continuum is measured for
the first time. Its presence is significantly required in all the
models adopted to describe the broadband emission, and in the PDS data
alone as well. The cut-off energy cannot be, however, very well constrained,
and values in the range 100-300~keV are possible.
A reanalysis of the 50--500~keV OSSE spectrum yields a steeper index
than measured by BeppoSAX (${\rm \Gamma_{OSSE} = 2.2 \pm^{0.2}_{0.3}}$),
and also the BATSE average spectrum shows evidence for a high-energy cutoff
above 100~keV (A.Malizia, private communication).
Such cut-off energies are well
consistent with those measured in several Seyfert~1s
so far, both by OSSE (Zdziarski et al. 1995; Madjeski et
al. 1995) and BeppoSAX (Piro et al.
1998; Guainazzi et al. 1999b; Guainazzi et al. 1999c; Perola et al. 1999).
Also the intrinsic
spectral index of the primary nuclear continuum (ranging between 1.4
and 1.7 in the various models) is slightly but not exceptionally flat
among Seyfert galaxies (Nandra et al. 1997; Turner et al. 1997).

\subsection{The 0.1--6~keV extended emission: unresolved discrete sources or truly diffuse emission?}

The total suppression of the transmitted component below $\simeq$8~keV
allows us to observe other spectral components, which are at least in part
associated with the host galaxy. The emission in the 0.1--6~keV energy
band is clearly extended along the plane of the host galaxy, thus
dismissing the possibility that the bulk of the intermediate X-ray
emission is due to scattering of the nuclear radiation only (Iwasawa
et al. 1993). In the MECS image at least two sources
are detected with a signal-to-noise ratio higher than 3. The spectral
properties and luminosity of the brightest are consistent with a
binary system accreting at sizeable fraction of the Eddington luminosity,
if it is indeed associated with NGC~4945.
Other explanations are, however, not excluded by our data. 
If the contribution of these sources is taken
into account, a significant extended emission remains within
at least the innermost 5~kpc, whose
extent is almost symmetric with respect to the nucleus. This would lead
to the conclusion that it is produced in a truly diffuse
interstellar medium. There is, however, no way of determining the
amount of the contribution of any unresolved underlying sources.
The ROSAT/PSPC images of NGC~4945 suggest that at least
50\% of the 0.1--2.4~keV photons are produced by discrete sources.
However, the extrapolation of the contribution of these sources in
the MECS band yields a fraction of the observed flux varying between
15\% and 50\%, if a power-law with ${\rm \Gamma}$ comprised between 1 and
2 is assumed.
The study of a large sample
of early type galaxies with ASCA and BeppoSAX suggests that the bulk of the
emission above 1~keV probably originates as the integrated emission
from X-ray binaries (Matsumoto et al. 1997; Trincheri et al. 1999).
A complete characterization
of the properties of this extended emission should await the
superior spatial resolution and sensitivity available with the scientific
payloads onboard {\it Chandra} and XMM.
What BeppoSAX data {\it can} clearly show is
that both the iron line and the continuum above 7~keV (where
the presence of the transmitted primary nuclear continuum starts
to be dominant) are, by contrast, produced by a point-like region
(at least as seen by the MECS). This strongly supports a nuclear origin
for both components.

\subsection{The broadband X-ray spectrum}

The 0.1--10~keV continuum can be fit with
the superposition of a single temperature thermal emission from a
collisionally excited plasma with ${\rm kT \simeq 3}$~keV and a warm
scattered nuclear power-law, with a scattering fraction $\simeq$0.4\%.
This model has already been successfully employed to describe the moderate
resolution intermediate X-ray spectrum of several Compton-thick Seyfert~2
galaxies
(Turner et al. 1997; Guainazzi et al. 1999a; Vignati et al.
1999). Two main differences are, however, present.
The warm scattered power-law
is seen through a substantial absorbing column density of $\simeq$$3 \times
10^{22}$~cm$^{-2}$. This might imply that the nuclear region is encompassed
by absorbing matter also on scales with are much larger than the dimension
of the Compton-thick structures, which almost suppress the direct view
of the nucleus. It is straightforward to associate it with the prominent
dust lanes crossing the galaxy plane, although the presence of a starburst
ring on a 100-pc scale (Marconi et al. 1999) provides a possible alternative
source of absorption (Fabian et al. 1999).
Recent HST NICMOS observations suggest that the active nucleus in NGC~4945
might be obscured along all lines of sight by matter with ${\rm N_H \sim
10^{21}}$--$10^{22}$~cm$^{-2}$ (Marconi et al. 1999).
On the other hand, the thermal
component has a much higher temperature than typically associated with the
X-ray emitting nuclear starburst in the Seyfert~2s
($\simeq$500~eV). Again, this may simply be telling us that the single
temperature description of this component is too a crude approximation.
Scenario a), however, can hardly be reconciled with the fact that the
bulk of the emission below 6~keV is extended on scales larger than
a few kpc. The warm scattered power-law,
in fact, dominates the fit in the whole 3--10~keV range,
therefore also where the emission is extended. This is almost at
odds with its supposed nuclear origin. The power-law could of course represent
only a phenomenological description of a different source of radiation,
{\it e.g.} the contribution of a population of unresolved discrete
sources.

A formally equivalent description of the soft/intermediate X-ray spectrum
is provided in the framework of the two thermal components template,
used to fit the integrated
bulge emission of X-ray faint early-type galaxies
(Bregman et al. 1995; Kim et al. 1996; Matsumoto et al. 1997;
Irwin \& Bregman 1999). The temperature
obtained with BeppoSAX for the hotter component
is remarkably consistent with
that measured on other early type galaxies, whereas the typical temperature
of the soft component is slightly lower than usual
(Matsumoto et al. 1997; Irwin \& Bregman 1999),
although consistent with that measured by ASCA in at least
one case (NGC~4392; Matsumoto et al. 1997). The fact that at least
50\% of the soft X-rays originate in discrete sources (see Sect.~3.2)
is in agreement with the results on M~31 (Primini et al. 1993; Irwin \&
Bregman 1999; Trincheri et al. 1999).
On the other hand,
Brandt et al. (1996) report a diffuse emission using the ROSAT/HRI on scales 
of 5$\arcmin$ along the plane of NGC~4945,
which is consistent with our reanalysis of the same data.
Obviously, there is no way
in this scenario to produce a neutral iron line with $>$1~keV EW. Actually,
such lines have never been observed in the integrated spectra of early
type galaxies (Matsushita et al. 1994; Matsumoto et al. 1997).
The origin
of the line must therefore be connected with the nuclear activity.

Recent studies suggest a possible alternative interpretation. In M~82
and NGC~253 (Cappi et al. 1999b), the intermediate X-rays are dominated
by hot gas with ${\rm kT \sim}$6--9~keV, probably associated with
starburst superwind outflows protruding from the galaxy disks. The observed
iron abundances are strongly sub-solar. If this scenario is viable
also for NGC~4945, these abundances can explain the lack of detection of
fluorescent emission from highly ionized iron. The presence of a wind-blown
cavity (Moorwood et al. 1996) strongly supports the existence of
a starburst superwind in NGC~4945. Again, high resolution imaging with
{\it Chandra} and XMM will provide invaluable contributions to resolve
this issue.

\subsection{The iron line: signature of the nuclear absorber?}

The spectral fits do not require the presence of any component associated
with Compton-reflection from the inner side of the putative molecular
torus surrounding the nuclear region, as observed in NGC~1068 (Matt et al.
1997) or Circinus~Galaxy (Matt et al. 1999a). This
rules out the most straightforward interpretation for the origin
of the prominent (${\rm EW \sim 1}$~keV) K$_{\alpha}$ fluorescent
iron line. The centroid energy and intrinsic narrowness imply an origin
from neutral or mildly ionized iron (Fe$<${\sc xx}). This is not in principle
inconsistent  with the line being produced in the same lukewarm
medium responsible, in scenario a), for the almost energy-independent
scattering of the primary nuclear radiation. However,
given also the problems of the scenario a), an appealing alternative
explanation is that the line is produced in transmission by
the same thick absorbing medium covering the nucleus
(Leahy et al. 1989).
Matt et al. (2000) compare the observed EW in the whole sample of the
Compton-thick Seyfert 2 galaxies observed by BeppoSAX with the values
expected in the pure transmission scenario, as calculated by 
the Matt et al. (1999b). In
NGC4945, the observed EW is consistent with the model predictions,
suggesting that the iron line is indeed due to transmission by the same
absorbing matter covering the nucleus. It is interesting to compare this
outcome with that for the 
Circinus Galaxy, where the observed value ($\simeq$60~eV
against the {\it total} continuum; Matt et al. 2000) is more than
one order of magnitude higher than expected from the column density of
the absorbing matter (${\rm 4 \times 10^{24}}$~cm$^{-2}$). In
Circinus~Galaxy a Compton reflection continuum is required to fit the
broadband X-ray spectrum (Matt et al. 1999a), most likely providing the
dominant contribution to the iron line. The difference between the
NGC4945 and Circinus~Galaxy case may be due to the geometry of the
absorbing matter, allowing in the latter the direct view of the farthest
side of the molecular torus encompassing the nucleus. The reader is
referred to Matt et al. (2000) for a more extensive discussion of this
issue.

\section{Conclusions}

The main results of this paper can
be summarized as follows:

\begin{itemize}
\item[a)] the 0.1-5~keV image profile is extended along the plane of the
host galaxy. The extension is symmetric within the innermost 5$\arcmin$, but
at larger radii only the SE MECS profile shows a broader tail. 
This asymmetry is due to two discrete
galactic sources which are present both in the soft X-ray ROSAT/PSPC
and in the intermediate X-ray MECS/BeppoSAX fields.
The profile perpendicular to the galaxy plane is
possibly also extended within $\pm$3$\arcmin$, although this result is made
more uncertain by possible residual inaccuracies in the BeppoSAX attitude
reconstruction
\item[b)] we measure the best high-energy ({\it i.e.}: $>$10~keV)
spectrum so far.  Along with the lack of significant extension in the
MECS emission above 7~keV, the spectral fitting
confirms the idea that the emission above 10~keV is dominated by
a nuclear non-thermal continuum, seen through an absorbing screen of
${\rm N_H \simeq}$ a few $10^{24}$~cm$^{-2}$. A flux variability with
a dynamical range $\simeq$60\% and extrapolated doubling/halving timescales
$<$~a few $10^4$~s (similar to that discovered
by Iwasawa et al. 1993 with {\it Ginga})
is {\it not} associated with any spectral changes,
suggesting that it echoes a, possibly more extreme, variability
of the intrinsic nuclear continuum
\item[c)] from our estimate of the intrinsic nuclear power, we conclude
that the energy output in NGC~4945 is likely to be
dominated by the AGN, not by
the starburst as suggested by infrared diagnostic (Genzel et al.
1998). NGC~4945 is therefore another indication that absorbed AGN
(and, possibly, also the so far elusive type~2 quasars) should be better
searched in X-rays rather in the optical/infrared band
\item[d)] the prominent iron line is consistent with K$_{\alpha}$ fluorescence
from neutral or mildly ionized iron ($<$Fe{\sc xx}). There is no evidence
for blending or intrinsic broadening at a level higher than 200~eV. 
The bulk of
the line (along with the continuum above 7~keV) is produced within about 1$\arcmin$,
strongly supporting a nuclear origin. The most likely explanation is
fluorescence in the same absorbing medium, responsible for the nuclear
absorption, provided its covering fraction is lower than one
\item[e)] the broadband 0.1--200~keV spectrum can be fitted with several
statistically equivalent models. However, the most consistent explanation
is in the framework of the two thermal components template, already
employed to fit the integrated bulge emission of early type galaxies,
where the bulk of the soft X-rays and a sizeable fraction of the
intermediate X-rays are believed to be produced by unresolved
discrete sources (mainly binaries). Alternatively, the extended emission
above 1~keV might be associated with a hot superwind-driven outflow, as
already observed in nearby starburst galaxies.
\end{itemize}

\begin{acknowledgements}

The BeppoSAX satellite is a joint Italian-Dutch program.
MG acknowledges an ESA Research Fellowship. WNB gratefully
acknowledges the support of NASA LTSA grant NAG5-8107.
GM acknowledges financial support from ASI and from MURST
(grant {\sc cofin}98--02--32).
The authors acknowledge
help from L.Chiappetti for the MECS data analysis, and useful comments
from A.Marconi and G.C.Perola. This research has made use of the NASA/IPAC
Extragalactic Database, which is operated by the Jet Propulsion Laboratory
under contract with NASA, and of data obtained through the High Energy
Astrophysics Science Archive Research Center Online Service, provided by
the NASA/Goddard Space Flight Center.

\end{acknowledgements}

\end{document}